# Full-color complex-amplitude vectorial holograms based on multi-freedom metasurfaces

Zi-Lan Deng[1,*,†], Mingke Jin[2,†], Xuan Ye[1], Shuai Wang[1], Tan Shi[1], Junhong Deng[2,3], Ningbin Mao[2], Yaoyu Cao[1], Bai-Ou Guan[1], Andrea Alù[4,5,6], Guixin Li[2,3,*], Xiangping Li[1,*]

[1]Guangdong Provincial Key Laboratory of Optical Fiber Sensing and Communications, Institute of Photonics Technology, Jinan University, Guangzhou 510632, China.

[2]Department of Materials Science and Engineering, Southern University of Science and Technology, 518055, Shenzhen, China.

[3]Shenzhen Institute for Quantum Science and Engineering, Southern University of Science and Technology, 518055, Shenzhen, China.

[4]Photonics Initiative, Advanced Science Research Center, City University of New York, 85 St. Nicholas Terrace, New York, NY 10031 USA

[5]Physics Program, The Graduate Center, City University of New York, 365 Fifth Avenue, New York, NY 10016 USA

[6]Department of Electrical Engineering, City College of New York, New York, NY 10031 USA

## Abstract

Phase, polarization, amplitude and frequency represent the basic dimensions of light, playing crucial roles for both fundamental light-mater interactions and all major optical applications. Metasurface emerges as a compact platform to manipulate these knobs, but previous metasurfaces have limited flexibility to simultaneous control them. Here, we introduce a multi-freedom metasurface that can simultaneously and independently modulate phase, polarization and amplitude in an analytical form, and further realize frequency multiplexing by a *k*-space engineering technique. The multi-freedom metasurface seamlessly combine geometric Pancharatnam-Berry phase and detour phase, both of which are frequency-independent. As a result, it allows complex-amplitude vectorial hologram at various frequencies based on the same design strategy, without sophisticated nanostructure searching of massive size parameters. Based on this principle, we experimentally demonstrate full-color complex-amplitude vectorial meta-holograms in the visible with a metal-insulator metal architecture, unlocking the long-sought full potential of advanced light field manipulation through ultrathin metasurfaces.

*E-mail: xiangpingli@jnu.edu.cn, ligx@sustech.edu.cn, zilandeng@jnu.edu.cn
† These authors contributed equally to this work.

Manipulating the optical wavefront over a multi-dimensional physical space, including phase, amplitude, polarization and frequency, not only paves the way for totally new light-matter interaction mechanisms, but also provides possibilities for various applications based on massive channels of optical multiplexing [1,2]. In this context, metasurfaces, or ultrathin functional layers, emerge as a desirable platform to manipulate the light field at will with large control and flexibility [3-7]. Exciting applications have already been demonstrated on the metasurface platform, including flat diffractive and polarization optical components, much more compact and lightweight than conventional bulky counterparts. By engineering the scattering properties of the individual meta-elements constituting the metasurface to mold the geometric phase, resonant phase or propagation phase, we are able to control phase [8,9], amplitude [10,11], polarization [12,13] or frequency [14-16] of light, leading to high-efficiency metalenses [9], high-fidelity holograms [8], broadband polarization components [12,17], and high-performance biosensors [15].

However, these ultrathin components tend to focus on single-dimensional light manipulation, controlling either the local phase, or amplitude, or polarization, or frequency, at a time, inherently limiting potential opportunities. For example, metasurface holograms and metalenses based on resonant phase and propagation phase are typically limited to a narrow range of frequencies [18, 19]. Geometric Pancharatnam–Berry (P-B) phase metasurfaces operate over broader bandwidths, but they are restricted to circular polarization only [8]. To improve the performance and enrich the functionality of metasurfaces for a broader range of applications, independent and simultaneous control of multiple dimensions of light over a single metasurface is highly desired. For example, simultaneous phase and amplitude control may be realized by complex meta-elements consisting of multiple geometric phase units to improve the quality of holographic image [20-22]. Complete phase and polarization control has also been achieved to perform vectorial wavefront-shaping [23-29]. Simultaneous control of both phase and frequency was also demonstrated, leading to advanced functionalities, such as vivid full-color holograms [30-33], broadband achromatic metalenses [34, 35], hybrid hologram color printing [36-38]. Nevertheless, all previous metasurfaces only access a limited number of dimensions of light manipulation [39, 40].

Different from resonant phase [18] and propagation phase [19], which rely on optimizing the size and shape of the meta-elements for a specific frequency of operation, the geometric P-B phase strategy [8] is totally decoupled from the spectral response, yielding an ideal broadband phase-modulation platform. Geometric P-B phase originates from the abrupt phase change when the spin angular momentum (SAM) of incident light transfers by going through a waveplate, as represented by the unit

Poincare sphere (Fig. 1a). Similarly, if the linear momentum (LM) of incident light transfers by going a grating, the light undergoes a *detour phase* [41-44]. In contrast to the Geometric P-B phase, which is proportional to the orientation angle of the meta-elements [45], the *detour phase* is proportional to the displacement between adjacent meta-elements. Both phases are wavelength independent and have a simply analytical expression with respect to the geometric parameters (orientation or displacement), without the need to search huge geometric parameter libraries for specified phase response.

Here, we propose a multi-freedom metasurface that seamless combines P-B phase and detour phase in a diatomic design, which is able to simultaneously control element-by-element the phase, amplitude and polarization of the impinging wavefront. Furthermore, the wavelength-independent phase nature and metagrating [46] architecture of the metasurface also enable frequency multiplexing through *k*-space engineering. Based on this principle, we experimentally demonstrated the multi-freedom metasurface based on metal-insulator-metal configurations in the visible frequency range, ultimately achieved full-color complex-amplitude vectorial meta-hologram and duplex polarization/color multiplexing.

The concept of generalized geometric phase is schematically illustrated in Fig. 1a. The initial and final states of light are located at the left and right Poincare spheres, whose centers represent the initial and final LMs $k_{x,0}$ and $k_{x,\sigma d}$, respectively. When the polarization state of light evolves from left-handed circular (LHC) (north pole of the Poincare sphere) to right-handed circular (RHC) (south pole) polarization, or vice versa, the geometric P-B phase is determined by the integral of the area swept by different longitudes over the Poincare sphere, where the longitudes represent evolution ways of polarization states when light passes through waveplates with different orientation $\psi$. At the same time, when the LM state of light evolves from the initial $k_{x,0}$ to the final $k_{x,\sigma_d} = k_{x,0} + \sigma_d 2\pi/p_0$, we can access the detour phase determined by the integral of the in-plane displacement $p$ with respect to the parallel wavevector $k_x$, $\varphi_d = \int_{k_{x,0}}^{k_{x,\sigma_d}} p\, dk_x = p\left(k_{x,\sigma_d} - k_{x,0}\right)$. As a result, the geometric P-B ($\varphi_p$) and detour ($\varphi_d$) phases have closed-form expressions with respect to the local orientation and displacement of the meta-element, $\varphi_p = \sigma_p 2\psi$ and $\varphi_d = \sigma_d 2\pi p / p_0$, respectively. Where $\psi$ and $p$ are the orientation angle and displacement respectively, and $\sigma_k$ ($k=p,d$) is an integer that quantifies the SAM/LM difference between final and initial states of light. For the P-B phase, $\sigma_p = \pm 1$, which represents half of the converted spin angular momentum from an LHC to an RHC beam, or vice versa. For the detour phase, $\sigma_d = \pm 1$, $\pm 2$, … represents the diffraction order of output light, which quantifies the linear momentum difference (in units of $2\pi/p_0$) between the diffraction and the incident beam. In a diatomic design with a proper fixed relative displacement between the meta-element pair, it can collaboratively fuse the two wavelength-independent geometric phases in higher-order diffraction states (Fig. 1b), providing massive degrees of freedom to control multiple dimensions of light simultaneously. By rigorous derivation (Supplementary Note 1), the Jones matrix dictated by four degrees of freedom $\psi_1$, $p_1$, $\psi_2$, and $p_2$ can be written as,

$$\mathbf{J} = iC \sin\psi_- \exp\left(i\pi \frac{p_+}{p_0}\right) \begin{pmatrix} -\sin\psi_+ & \cos\psi_+ \\ \cos\psi_+ & \sin\psi_+ \end{pmatrix}, \tag{1}$$

where, $p_+=p_1+p_2$, $\psi_\pm=\psi_1\pm\psi_2$. For circularly polarized incidence $\mathbf{E}_i^{CP} = (1 \ \ i)^T$, the output field is $\mathbf{E}_o^{CP} = iC\sin\psi_- e^{i\pi p_+/p_0} e^{i\psi_+}(i \ \ 1)^T$, where the geometric P-B phase term $e^{i\psi_+}$ and detour phase term $e^{i\pi p_+/p_0}$ are decoupled from each other. For linearly polarized incidence, e.g., transverse electric (TE) polarization $\mathbf{E}_i = (0 \ \ 1)^T$, the output field becomes

$$\mathbf{E}_o = \mathbf{J}\begin{pmatrix}0\\1\end{pmatrix} = iCe^{i\delta_1}\sin\psi_-\begin{pmatrix}\cos\psi_+\\\sin\psi_+\end{pmatrix}. \tag{2}$$

Closed-form expressions of the phase $e^{i\delta_1}$, amplitude $\sin\psi_-$, and polarization orientation $(\cos\psi_+, \ \sin\psi_+)^T$ of the impinging wavefront are analytically derived in terms of global displacement $p_+$, orientation angle difference $\psi_-$, and orientation angle sum $\psi_+$, respectively.

As the both the P-B phase and detour phase are completely decoupled from the spectral response by using uniformly-sized meta-elements, simultaneously modulated phase, amplitude and polarization are applicable to broadband light in this design approach. Intriguingly, this platform can be further extended to frequency multiplexing of light, assuming that multiple frequencies are carrying different images. In order to achieve this goal, we apply *k*-space engineering to split the different colored hologram images in different spatial locations and construct the full-color image in a target area using grating dispersion (Fig. 1c), which is much convenient than previous approaches requiring multiple incident angles [31, 32, 47].

The multi-freedom metasurface works in the extraordinary optical diffraction (EOD) regime, where only the 0[th] and -1[st] diffraction orders are allowed for propagating waves. To tune the EOD resonance at visible frequencies, we use aluminum nanorods (length $L$=180 nm, width $w$=80 nm, periodicity $p_x$=$p_0$=500 nm, $p_y$=300 nm) on top of a $SiO_2$ spacer (height $h$=100 nm) and an aluminum background film (Fig. 2a). The diffraction efficiency of such aluminum nanorod arrays is significantly enhanced at broadband visible frequencies and wide angles (Fig. 2b). Due to the intrinsic Ohmic loss of metal (green curve in Fig. 2c), the diffraction efficiency reaches 70%, while the $0^{th}$ diffraction efficiency is effectively suppressed to near zero (red curve in Fig. 2c). The high diffraction efficiency can be obtained in the visible range because the interband absorption peak of aluminum is located at longer wavelengths (green curve in Fig. 2c). The EOD resonance is highly anisotropic, i.e., the diffraction efficiency for the transverse-magnetic (TM) polarization (cyan dashed curve in Fig. 2c) is much lower than for TE polarization, which forms the basis for the multi-freedom metasurface construction in a diatomic design. We note that there is a sharp absorption resonance peak near 500 nm for the TM polarization, which is due to the excitation of propagating surface plasmon modes in the $x$ direction. The amplitude modulation $\sqrt{R_{-1}}$ by a meta-molecule formed by an aluminum nanorod-pair is shown in Figs. 2d-f. Here, the parameters of each nanorods are $w$ = 60 nm, $L$ = 140 nm, $p_0$ = 500 nm, $p_y$ = 350 nm, optimized to ensure sufficient extinction ratios between orthogonal polarizations. The modulated amplitude generally goes from zero to the maximum when the orientation angle difference $\psi_-$ of the two meta-elements changes from 0º (180º) to 90º, although with a slight deviation from the ideal sine function predicted by Eq. (2). In the short wavelength range, the amplitude maximum is slightly shifted from $\psi_-$=90º, due to the reduced diffraction extinction ratio of nanorods in the short wavelength range. The modulation of polarization state $tan\Psi \cdot e^{i\delta}$ by $\psi_+$ is shown in Figs. 2g-i. The amplitude ratio angle $\Psi$ between the two orthogonal field components is a linear function of $\psi_+$ (Fig. 2h), and the phase difference $\delta$ between the two orthogonal components is constant at $\pi$ (Fig. 2i). These relationships span over the entire visible frequency range, although there is a small bump near 550 nm due to unwanted excitation of a propagating surface plasmon mode. Besides the amplitude and polarization state modulations, the control over the phase profile is based on the detour phase scheme [44] determined by an independent degree of freedom, namely the displacement sum $p_+$ [41-43]. Therefore, the aluminum meta-molecule geometry introduced here is indeed capable of complete phase, amplitude and polarization control in the whole visible light range.

Since the multi-freedom metasurface works directly in the -1$^{st}$ diffraction order rather than the 0$^{th}$ diffraction order, it offers the flexibility to realize frequency multiplexing by $k$-spacing engineering using grating dispersion of the metagrating architecture. When a white light composed of three primary colors (*R*: 671 nm, *G*: 561 nm, *B*: 473 nm) illuminates the metasurface with incident angle $\theta_0$ (Fig. 3a), the diffraction angles of different frequencies will be largely separated (Fig. 3b). The output beams ($k_{-1}$) for different frequencies have a fixed momentum difference $2\pi/p_0$ compared with the incident light ($\theta_0$), and lead to different deflection angles $\theta_R, \theta_G, \theta_B$, determined by both the parallel wavevector ($k_x$) and the wavenumber $k_{0J}$ (*J*=*R*,*G*,*B*) as shown in Fig. 3b. The diffracted angles for *R, G, B* components can be readily obtained as follows (Supplementary Fig. S2 and Supplementary Note 2),

$$\sin\theta_J = \lambda_J / p_0 - \sin\theta_0, \tag{3}$$

where $J=R, G, B$ represents the three primary color components. The observation plane is set to be perpendicular to the diffraction direction of the central wavelength (G: 561nm), and the distance between the observation plane and the multi-freedom metasurface is $z_d=z_{dG}$. The other two components ($R$, and $B$) of diffracted light will be automatically deflected from the G component with separation angles ($\delta_R=\theta_R-\theta_G$, $\delta_B=\theta_G-\theta_B$) and propagation distances ($z_{dR}=z_d/\cos\delta_R$, $z_{dB}=z_d/\cos\delta_B$), respectively. To obtain the pre-designed full-color holographic image in the target observation zone (dashed frames in Fig. 3c), we can add phase-shift factors $\Delta\varphi_R=k_R x \sin\delta_R$, $\Delta\varphi_G=0$, $\Delta\varphi_B=-k_B x \sin\delta_B$, to complex-amplitude holograms of the $R, G, B$ components, respectively. These phase-shift factors are automatically introduced by the momentum shift associated with the diffraction of different wavelengths, in stark contrast with the input momentum shift adopted in previous full-color holographic approaches based on conventional techniques[48] and metasurfaces [31, 32, 47]. As a result, the complex-amplitude hologram summing the three components can be written as,

$$\tilde{A}_{tot}(x,y) = \sum_{J=R,G,B} A_J(x,y) e^{i\varphi_J(x,y)+i\Delta\varphi_J}, \qquad (4)$$

where, $A_J$ and $\varphi_J$ ($J=R, G, B$) are the amplitude and phase profiles of $R, G, B$ components, respectively.

By illuminating the multi-freedom metasurface separately with $R, G, B$ laser beams, the total reconstructed image profiles will be (see Supplementary Note 2),

$$I_R^{re} \sim I_R(x_i, y_i) + I_G(x_i - Z_d \tan\delta_R, y_i) + I_B(x_i - Z_d \tan\delta_B - z_d \tan\delta_R, y_i), \qquad (5a)$$

$$I_G^{re} \sim I_G(x, y) + I_R(x - z_d \tan\delta_R, y) + I_B(x - z_d \tan\delta_B, y), \qquad (5b)$$

$$I_B^{re} \sim I_B(x, y) + I_R(x + z_d \tan\delta_R + z_d \tan\delta_B, y) + I_G(x + z_d \tan\delta_B, y), \qquad (5c)$$

respectively. Where $I_J^{re}$ and $I_J$ (J=R, G, B) denote the intensity profiles of the reconstructed and original image of color $J$, respectively. As we can see, the recorded image profile for $R$, $G$, $B$ components can be precisely reconstructed as the first term of the right hand side (RHS) in Eqs. (5a-c), when the correct input beam is used. At the same time, there are unwanted crosstalk images [the second and third terms of RHS in Eqs. (5a-c)]. Nevertheless, those crosstalk images appear outside of the target observation zone of the holographic image, which is predesigned outside the maximum allowed numerical aperture NA=sin(min($\delta_R$, $\delta_B$)/2) (See Supplementary Note 2 and Table S1). Compared with previous full-color meta-hologram schemes [30-33], the multi-freedom metasurface platform significantly relieves both optical setup and metasurface structures (see comparison in Supplementary Table S2).

The theoretical and experimental results of holographic images separately reconstructed by three primary color beams based on this full-color holographic approach are shown in Fig. 3c. For each primary color, three separate images are reconstructed from all holographic information channels. In the central area of the observation plane (enclosed by the dashed yellow frame), the correct holographic images in the corresponding color channel appear clear and undistorted, with the same size, while the crosstalk images that stem from additional shifted terms in Eqs. (5a-c) are distorted and the sizes appear enlarged (shrunk) for R (B) components, due to the scaling factor between size and wavelength[48]. The comparison between phase-only and complex-amplitude full-color meta-holograms is shown in Fig. 3d. The phase-only color-hologram is constructed using detour phase with a single meta-element per unit cell (right-up panel of Fig. 3c), while the complex-amplitude color-hologram uses the generalized geometric phase with a diatomic design (right-down panel of Fig. 3d). Speckles are visible in the phase-only holograms, due to the requirement of introduction of random phase and multiple iteration steps to homogenize the amplitude distribution in the hologram calculation. For the complex-amplitude hologram, the full-color holographic image is much smoother and clearer, as both phase and amplitude information of the wavefront are recovered by the metasurface.

To demonstrate the vectorial properties of the full-color meta-hologram, we designed two sets of holograms that can encode two full-color images with orthogonal polarization orientations (Fig. 4) by further exploiting the degree of freedom $\psi_+$. $\psi_+$ is fixed at 90º for the first set of meta-molecules, while $\psi_+=45º$ for the second hologram. Therefore, the first full-color image (*rainbow*) has the same polarization as the incident light, while the second full-color image (*flower*) has a polarization rotated $2\psi_+=90º$ with respect to the incident light. In the experiment, a broadband polarizer-analyzer pair was used before and after the metasurface. Fixing either one of the polarization states and flipping the other polarization state, the appearance of the full-color image will change from one image to the other. Based on the multi-freedom metasurface, we also demonstrated a multi-channel multiplexed hologram (Supplementary Fig. S6) with polarization-pair dimensions containing three independent channels (vertical/vertical, vertical/horizontal, horizontal/horizontal) and the color dimension containing three channels (R, G, B), which significantly increases the information capacity, as well as information security in multiplex-oriented applications.

**Discussion**

The proposed multi-freedom metasurface provides unprecedented flexibility for wavefront engineering of light. It can independently and simultaneously control phase, amplitude, polarization of light, and at the same time perform the frequency multiplexing, all of which are achieved only with a simply single-sized meta-atom design and at a single ultrathin layer. These properties are highly desirable in practical metasurface devices.

Although we have demonstrated the multi-freedom metasurface using plasmonic nanorods in reflection mode, the concept can be straightforwardly extended to high-efficiency all-dielectric transmission metasurfaces. The multi-freedom design principle unlocks the long-sought full potential of metasurfaces, and underpins fascinating applications in high-fidelity full-color holographic displays, quantum entanglements and exotic vectorial field reconstructions.

**Methods**

**Simulation design of the multi-freedom metasurface.** The reflection amplitude and phase spectra at multiple diffraction orders (mainly the $0^{th}$ and $-1^{st}$ diffraction orders) of the aluminum nanorod array were calculated using FEM methods implemented by COMSOL. The construction of the full-color hologram was first calculated by the Fresnel diffraction formulas for each primary color component, and then was combined together with proper phase shift terms according to a k-spacing engineering technique.

**Fabrication and measurement of the multi-freedom metasurface**. The metasurfaces were fabricated on a silicon substrate. First, we used electron-beam evaporation to deposit a layer of 130-nm-thick aluminum film and a layer of 100-nm-thick $SiO_2$ film, successively. Next, the 30 nm-thick aluminum nanorod arrays were patterned on the $SiO_2$ film by the Electron-beam lithography followed by a lift-off process. To construct the vectorial full-color holographic image, three continuous-wave laser beams at wavelengths of 473 nm, 561 nm and 671 nm are combined to generate the white light source to illuminate the metasurface from the same direction. The diffracted full-color image was displayed on a white screen, and captured by a Nikon Camera.

**Acknowledgements**

This work was supported by National Key R&D Program of China (YS2018YFB110012), National Natural Science Foundation of China (NSFC) (Grant 11604217, 61522504, 61420106014, 11774145, 11734012, 11574218, 51601119), the Guangdong Provincial Innovation and Entrepreneurship Project (Grant 2016ZT06D081, 2017ZT07C071), the National Science Foundation, the Applied Science and Technology Project of the Guangdong Science and Technology Department (2017B090918001) and the Natural Science Foundation of the Shenzhen Innovation Committee (JCYJ20170412153113701). Z.-L.D. is also funded by the China Scholarship Council.

**Author contributions**

Z.-L.D. initiated the idea. Z.-L.D., X.L. and G.L. designed the experiments. Z.-L.D carried out the design, theoretical analysis and simulation of the metasurfaces. M.J.,

J.D., N.M., and G.L. fabricated the samples. Z.-L.D, X.Y., T.S., and S.W. performed the measurements. Z.-L.D, G.L, and X.L analyzed the data. Z.-L.D, X.L., G.L., and A.A. wrote the manuscript . All authors contributed to discussions about the manuscript.

## Conflict of interests

The authors declare no conflicts of interest.

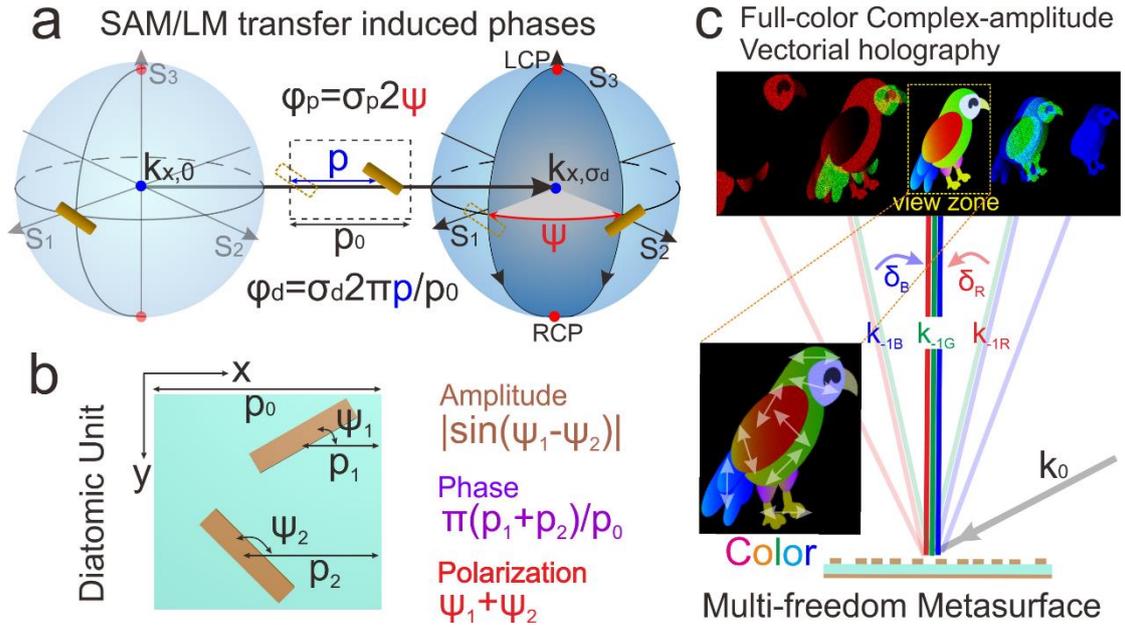

**Fig. 1. Design principle of the multi-freedom metasurfaces.** (a) Two kinds of frequency-independent phase modulations with SAM/LM transfer. The geometric P-B phase $\varphi_p$ is induced by SAM transfer from LCP to RCP or vice versa, represented by the Poincare sphere, depending only on orientation angle $\psi$, and the detour phase $\varphi_d$ is induced by LM transfer, depending only on displacement $d$. (b) Diatomic meta-molecule design that fuses the P-B and detour phases with freely controllable orientations ($\psi_1, \psi_2$) and displacements ($p_1, p_2$) (left). It can modulate the amplitude, phase and polarization of the local field simultaneously by three closed-form expressions: amplitude $A=|\sin(\psi_1-\psi_2)|$, phase $\varphi=\pi(p_1+p_2)/p_0$, and polarization angle $\psi=\psi_1+\psi_2$ (right). (c) Schematic illustration of the overall multi-freedom metasurfaces and the realization of frequency multiplexing by a *k*-space engineering technique, leading to a full-color complex-amplitude vectorial meta-hologram.

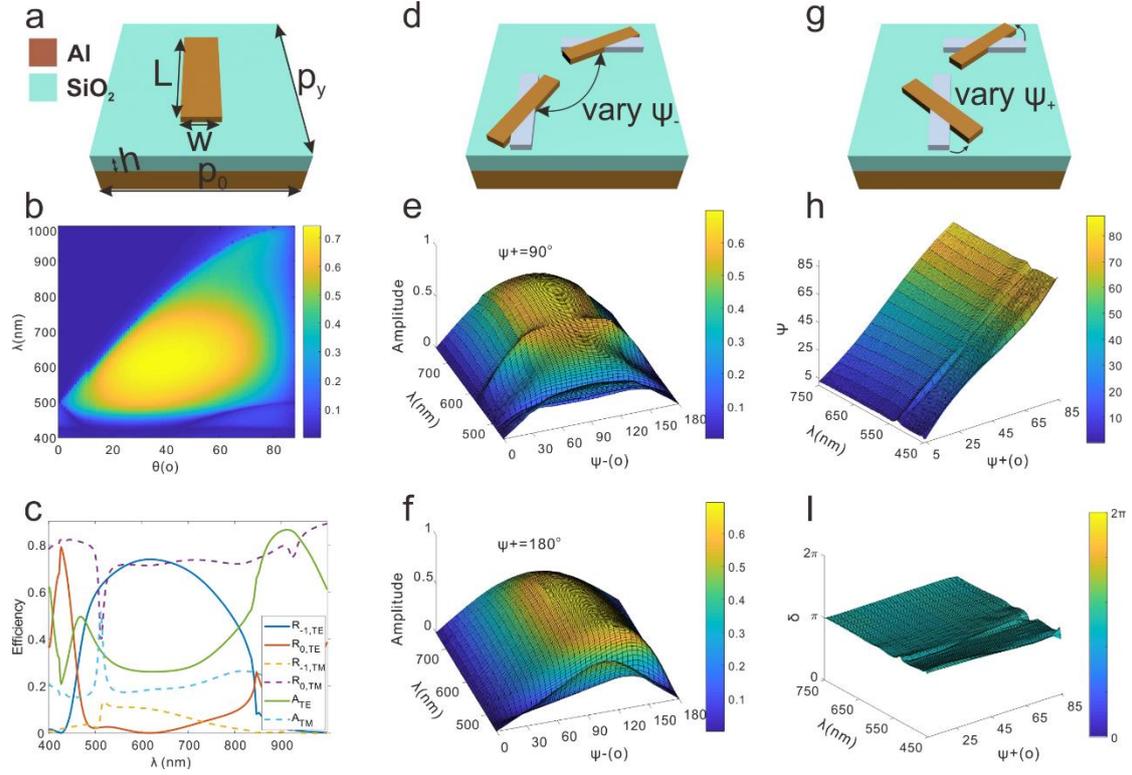

**Fig. 2. Diffraction and modulation properties of the periodic multi-freedom metasurface in visible frequencies.** (a) Illustration of one unit-cell of the plasmonic nanorod array with EOD in visible frequencies. (b) Enhanced diffraction efficiencies ($R_{-1, TE}$) of the aluminum nanorod array, for different incident angles and wavelengths. (c) Diffraction (-1$^{st}$ order) and absorption spectra under both transverse electric (TE) and transverse magnetic (TM) illumination (incident angle 45º) of the periodic aluminum nanorod array in the MIM configuration with length L=180 nm, width w=80 nm, thickness t=30 nm and dielectric spacer height h=100 nm. (d-f) The modulated amplitude profile as a function of orientation angle difference $\psi_-$ in the visible range, at orientation angle sum (e) $\psi_+=90°$, and (f) $\psi_+=180°$, respectively. (g-i) The modulated polarization state ($\tan\Psi e^{i\delta}$) parameters: (h) polarization amplitude ratio $\Psi$ and (i) relative phase difference $\delta$ as a function of orientation angle sum $\psi_+$, respectively.

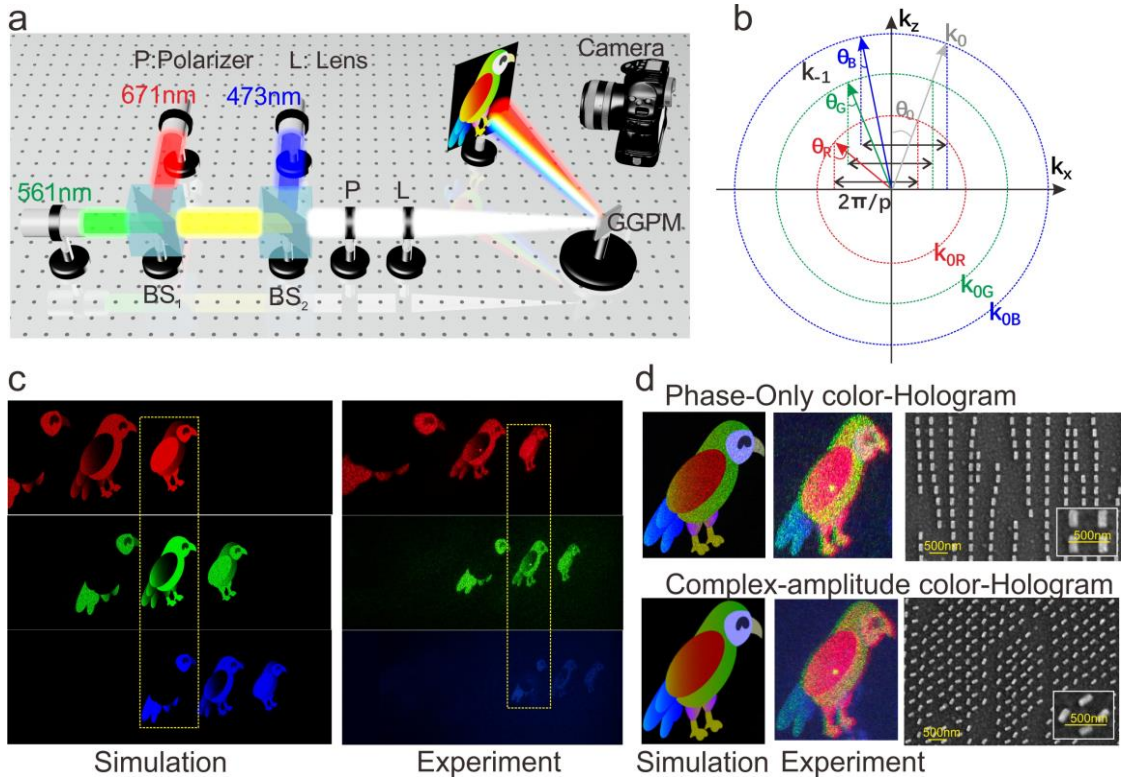

**Fig. 3. Experimental realization of multi-freedom metasurface**. (a) Experiment setup for the full-color hologram reconstruction. (b) Wavevector diagram illustrating the relation between output ($k_{-1}$) and incident light ($k_0$), which has a fixed momentum difference $2\pi/p_0$, the incident angle and three diffraction angles are $\theta_0$, $\theta_R$, $\theta_G$, $\theta_B$, respectively. (c) The overall reconstructed holographic images separately by the R (671 nm), G (561 nm), and B (473 nm) CW beams. The areas surrounded by the dashed lines are the target zone that forms the predesigned full-color images. Left and right panels show the simulation and experimental results, respectively. (d) The reconstructed full-color holographic images by illuminating the metasurface with white light formed by the R, G, B beams with the same incident direction. The upper panels and the lower panels show the case of phase-only and the complex-amplitude full-color holograms, respectively. The SEM images showcase fabricated metasurfaces composed of aluminum nanorod arrays for phase-only (upper panel) and amplitude phase full-color holograms (lower panel), respectively.

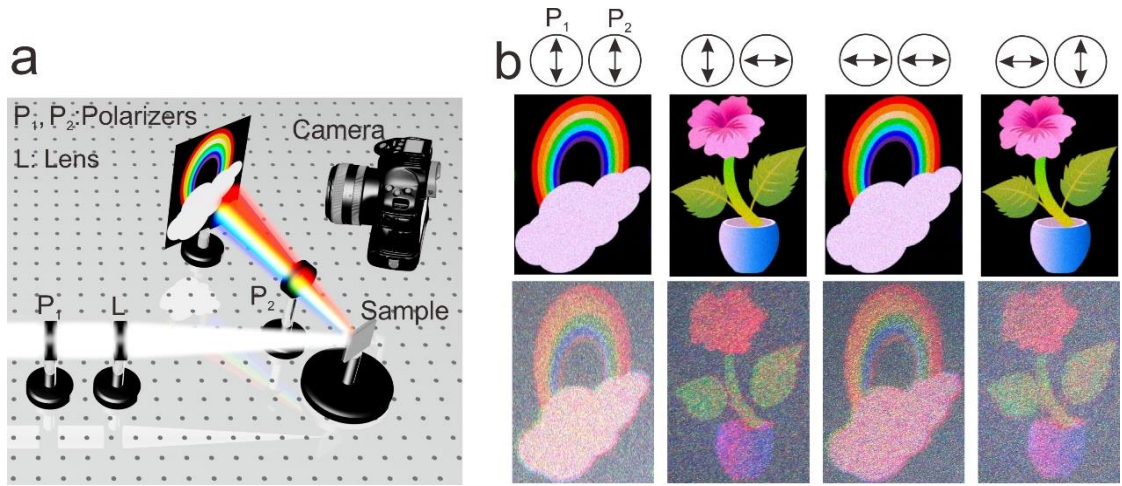

**Fig. 4. Mutli-freedom metasurface for double-way polarization switched full-color vectorial holographic images** (a) Experimental setup for the double-way holographic switching, two polarizers P$_1$ and P$_2$ are located before and after the metasurface. The combination of polarization orientations of those two polarizers determines the appearance of different holographic images. (b) The simulated (upper panel) and experimental results (lower panel) of reconstructed full-color images for different combinations of polarization orientations. The full-color *rainbow* and *flower* images alternatively appear when we fix the polarization switch 1 (P$_1$) while flipping the polarization switch 2 (P$_2$); or fix switch 2 while flipping switch 1.